\documentclass[a4paper,12pt,titlepage]{article}

\usepackage{amsmath,amssymb}
\usepackage{geometry}
\usepackage{hyperref}
\usepackage{braket}
\usepackage{titlesec}
\usepackage{fancyhdr}
\usepackage[dvipdfmx]{graphicx}
\usepackage{slashed}
\usepackage{amsthm}
\usepackage{cite}

\theoremstyle{definition}
\newtheorem{theo}{Theorem}[section]
\newtheorem{defi}[theo]{Definition}

\hypersetup{
    setpagesize=false,
    bookmarksnumbered=true,
    bookmarksopen=true,
    colorlinks=true,
    linkcolor=blue,
    citecolor=blue,
}

\begin{document}

\fancypagestyle{foot}
{
    \fancyfoot[L]{\(^{*}\)E-mail address: yakkuru$\_$111@ruri.waseda.jp}
    \fancyfoot[C]{}
    \fancyfoot[R]{}
    \renewcommand{\headrulewidth}{0pt}
    \renewcommand{\footrulewidth}{0.5pt}
}

\renewcommand{\footnoterule}{%
  \kern -3pt
  \hrule width \columnwidth
  \kern 2.6pt}


\begin{titlepage}
    \begin{flushright}
        \begin{minipage}{0.2\linewidth}
            \normalsize
            WU-HEP-20-08
        \end{minipage}
    \end{flushright}

    \begin{center}
        \vspace*{5truemm}
        \Large
        \bigskip \bigskip
        \LARGE \textbf{Magnetized Riemann Surface of Higher Genus and Eta Quotients of Semiprime Level} 
        
        \Large

        \bigskip \bigskip
        Masaki Honda$^{1,*}$ 

        \smallskip

        {\large $^{1}$ {\it Department of Physics, Waseda University, Tokyo 169-8555, Japan}} \\ 

        \bigskip \bigskip \bigskip
        
        \large \textbf{Abstract}
    \end{center}
    We study the zero mode solutions of a Dirac operator on a magnetized Riemann surface of higher genus. In this paper, we define a Riemann surface of higher genus as a quotient manifold of the Poincar$\acute{\text{e}}$ upper half-plane by a congruence subgroup, especially $\Gamma_{0}(N)$. We present a method to construct basis of cusp forms since the zero mode solutions should be cusp forms. To confirm our method, we select a congruence subgroup of semiprime level and show the demonstration to some lower weights. In addition, we discuss Yukawa couplings and matrix regularization as applications.
    \thispagestyle{foot}
    
\end{titlepage}

\baselineskip 7.5mm


\tableofcontents

\section{Introduction}
\label{1}
The standard model (SM) is a successful model to explain the results of high energy experiments. However, there are several problems to be solved, e.g., the origin of the generations and the chiral structure. Combining recent cosmological observations, they imply the existence of a new theory beyond the SM.

Superstring theory is a candidate of a unified theory of all forces in nature. This theory requires a ten dimensional (10D) space-time because of theoretical consistency. Although there are several works for 10D space-time, we focus on extra dimensional models and matrix models in this paper.

In extra dimensional models, we attempt to the problems of the SM from geometry and analysis of an extra dimensional space. Typical method is spectral decomposition, e.g., Kaluza-Klein dimensional reduction. In four dimensional (4D) effective theories, the eigenvalues contribute to mass parameters and the eigenfunctions can induce the generations and the chiral structure. In addition, overlap integrals, which is a integral of products of the eigenfunctions over the extra dimensional space, contribute some parameters like Yukawa couplings. From the phenomenological point of view, a Dirac operator is selected as a target of the spectral decomposition. The zero mode solutions of the Dirac equation are important since fermions in the SM are massless before the spontaneous symmetry breaking,

On the other hand, superstring theory is established at only perturbative level. Accordingly, it has a lot of perturbatively stable vacua. Therefore, it implies the necessity of a fundamental theory. Matrix models are proposed as nonperturbative formulations of superstring theory. In matrix models, the space-time does not exist a priori since information of the space-time is embedded into matrices and the matrices follow dynamics. Therefore, the dynamics of the matrices determines the space-time including an extra dimensional space. In matrix models, a fundamental process is matrix regularization. The matrix regularization is an approximation of a Poisson algebra by a matrix algebra. The original Poisson algebra recovers in the limit of infinite matrix size. One of the methods is to use the zero mode solutions of a Dirac operator~\cite{Bordemann:1993zv,Ma:2008,Ishiki:2018aja}. The zero mode solutions of a Dirac operator in a $U(1)$ gauge background are known to have a graded ring structure whose grade is characterized by the value of the magnetic flux. Based on this fact, the authors of Ref.~\cite{Adachi:2020asg} showed the matrix regularization for sphere and torus.

In both cases, the problem is how to derive the zero mode system (degeneracy, chirality and functional form) of a Dirac operator in a $U(1)$ gauge background. Fortunately, degeneracy and chirality can be derived from the index theorem. However, there are few examples in which functional forms are specifically derived. There are only magnetized\footnote{We use ``magnetized'' to describe that something is in a $U(1)$ gauge background} sphere~\cite{Conlon:2008qi} and magnetized torus~\cite{Cremades:2004wa} for closed oriented two dimensional manifold. Therefore, the question of whether a zero mode system can be constructed for Riemann surfaces of higher genus (genus 2 or more) naturally arises.

In Ref.~\cite{Adachi:2020asg}, Riemann surfaces of higher genus are realized as quotient manifolds of the Poincar$\acute{\text{e}}$ Disk by some discrete groups. The authors showed that zero mode solutions on a magnetized Riemann surface of higher genus are automorphic forms. A set of automorphic forms becomes a vector space. This leads us to construct the basis of such a vector space.

The purpose of this paper is explicit construction of such a vector space by selecting a discrete group.

The characteristics of automorphic form do not change even in the Poincar$\acute{\text{e}}$ upper half-plane, which is conformally equivalent with the the Poincar$\acute{\text{e}}$ Disk. In addition, Riemann surfaces of higher genus is also realized as quotient manifolds of the Poincar$\acute{\text{e}}$ upper half-plane by some discrete groups. Moreover, we can apply useful formulae. Therefore, in the following, we consider the Poincar$\acute{\text{e}}$ upper half-plane and its quotient manifolds.

In this paper, we adopt eta quotients to construct basis of zero mode solutions. Eta quotients are defined by products of the Dedekind eta function. In Ref.~\cite{Rouse:2013ar,Allen:2019ar}, eta quotients are used to describe the basis of automorphic forms and elliptic curves, respectively. We apply their method to describe the basis of cusp forms, which are automorphic forms satisfying an additional condition. Moreover, to confirm our method, we demonstrate for cusp forms of weight 2,4 and 6 by selecting the discrete subgroup is $\Gamma_{0}(35)$. 

The organization of this paper is as follows. In Section \ref{2}, we briefly review the Poincar$\acute{\text{e}}$ upper half-plane, automorphic forms, cusp forms and eta quotients. In Section \ref{3}, we construct a Dirac operator on a magnetized Riemann surface of higher genus based on Ref.~\cite{Adachi:2020asg} with some modifications. Additionally, we consider conditions for the convergence and the chirality of zero mode solutions. In section \ref{4}, we construct zero mode solutions from eta quotients. We will show a demonstration for cusp forms of weight 2,4 and 6 by selecting the discrete subgroup is $\Gamma_{0}(35)$. In section \ref{5}, we apply our result to extra dimensional models and matrix regularization. We will discuss Yukawa couplings and matrix regularization of Riemann surfaces of higher genus. Section \ref{6} contains conclusions and discussion.

\section{Preliminaries}
\label{2}

\subsection{Poincar$\acute{\text{e}}$ upper half-plane}
\label{2.1}
The Poincar$\acute{\text{e}}$ upper half-plane

\noindent
\begin{align*}
    \mathbb{H} := \{ z=x+iy \mid x,y \in \mathbb{R}, y>0 \}
\end{align*}
is a Riemann manifold with the metric

\noindent
\begin{align}
\label{metric}
    ds^{2}=y^{-2}(dx^{2}+dy^{2}).
\end{align}

To describe Riemann surface of higher genus, we consider well-known properties of the M$\Ddot{\text{o}}$bius transformations as group actions of $SL(2,\mathbb{R})$

\noindent
\begin{align}
\label{Mobius}
    \gamma=
    \begin{pmatrix}
    a & b \\
    c & d
    \end{pmatrix} \in SL(2,\mathbb{R}), \quad
    z \mapsto \gamma(z) := \frac{az+b}{cz+d}.
\end{align}

A discrete subgroup $\Gamma$ of $SL(2,\mathbb{R})$ is called first kind Fuchsian group if every point on $\partial \mathbb{H} = \mathbb{R} $ is a limit of the orbit $\Gamma z$ for some $z \in \mathbb{H} $. For a first kind Fuchsian group $\Gamma$, $\gamma \in \Gamma$ is called parabolic if $\gamma \neq \pm \mathbf{1} $ and the eigenvalues are $1$ only or $-1$ only. For an parabolic element $\gamma \in \Gamma$, a point $z \in \mathbb{R} \cup \{\infty \}$ satisfying $\gamma(z)=z$ is uniquely determined, and such a point is called cusp.

Two points $z,z' \in \mathbb{H}$ are equivalent if $z' \in \Gamma z$. This is an equivalent relation and naturally leads to consider a quotient $\Gamma \backslash \mathbb{H}$. In general, the quotient $\Gamma \backslash \mathbb{H}$ is not compact. To obtain a compact manifold, we consider the extended upper half-plane $\Tilde{\mathbb{H}}:=\mathbb{H} \cup \mathbb{Q} \cup \{\infty \} $. By restricting ourselves to $SL(2,\mathbb{Z})$ and its discrete subgroups, the M$\ddot{\text{o}}$bius transformations (\ref{Mobius}) can be extended to $\Tilde{\mathbb{H}}$, and a quotient $\Gamma \backslash \Tilde{\mathbb{H}}$ (this is equivalent to $\Gamma \backslash \mathbb{H}$ with cusps) is compact. From this fact and the result of Poincar$\acute{\text{e}}$-Koebe's uniformization theorem, any Riemann surface can be written as $\Gamma \backslash \Tilde{\mathbb{H}}$ for some Fuchsian group $\Gamma$ and hence inherits a hyperbolic metric (with singularities if $\Gamma$ has fixed points).

In the following, we select a congruence subgroup $\Gamma_{0}(N)$ as a Fuchsian group $\Gamma$. A congruence subgroup of level $N$, $\Gamma_{0}(N)$, is defined by

\noindent
\begin{align*}
    \Gamma_{0}(N):= \left\{ \gamma \in SL(2,\mathbb{Z}) \mid \gamma=\begin{pmatrix}
    a & b \\
    c & d
    \end{pmatrix}
    \equiv 
    \begin{pmatrix}
    \ast & \ast \\
    0 & \ast
    \end{pmatrix}
    (\operatorname{mod}\ N)
    \right\}
\end{align*}
where $N$ is a positive integer and $\ast$ means no condition, except $\operatorname{mod}\ N$, on this element.

\subsection{Automorphic forms and cusp forms}
\label{2.2}

The M$\ddot{\text{o}}$bius transformations~(\ref{Mobius}) induce the action of $SL(2,\mathbb{R})$ and its subgroups on functions on $\Tilde{\mathbb{H}}$.

\noindent
\begin{defi}
\label{def forms}
    Let $\gamma=
    \begin{pmatrix}
    a & b \\
    c & d
    \end{pmatrix}
    \in \Gamma_{0}(N)$,
    $k$ be a positive integer and $f:\mathbb{H} \to \mathbb{C}$. The function $f$ is an automorphic form of weight $k$ with respect to $\Gamma_{0}(N)$ if $f$ satisfies the following conditions.
    \begin{enumerate}
    \renewcommand{\labelenumi}{(\roman{enumi})}
        \item $f$ is holomorphic on $\mathbb{H}$,
        \item $f$ is holomorphic on cusps of $\Gamma_{0}(N)$,
        \item $f \left( \gamma(z) \right) = (cz+d)^{k} f(z), \quad  ^{\forall}\gamma \in \Gamma_{0}(N)$.
    \end{enumerate}
    If a function $f$ is an automorphic form and $f=0$ at all cusps of $\Gamma_{0}(N)$, the function $f$ is called cusp form.
\end{defi}

We note that there are some extensions: the weight can be an half integer, and the transformation law (iii) can have some character $v(\gamma)$ such that $|v(\gamma)|=1$.

Let $M_{k} \left( \Gamma_{0}(N) \right)$ and $S_{k} \left( \Gamma_{0}(N) \right)$ be sets of automorphic forms and cusp forms, respectively. From the definition~\ref{def forms} and $\Gamma \backslash \Tilde{\mathbb{H}}$ is compact, they are finite dimensional vector spaces on $\mathbb{C}$. We can calculate their dimensions based on Riemann-Roch theorem.

\begin{theo}
If $k \geq 2$ is even, the dimensions of $M_{k}\left( \Gamma_{0}(N) \right)$ and $S_{k} \left( \Gamma_{0}(N) \right)$ are given by
\begin{align}
    &\operatorname{dim}M_{k}(\Gamma_{0}(N),v)= (k-1)(g-1) + \frac{k}{2} v_{\infty} + \left[\frac{k}{4} \right] v_{2} + \left[ \frac{k}{3} \right]v_{3},\notag \\
    \label{casp-dimformula}
    &\operatorname{dim}S_{k}(\Gamma_{0}(N),v)=\operatorname{dim}M_{k}(\Gamma_{0}(N),v)-v_{\infty} + \delta_{k,2},
\end{align}
\end{theo}
\noindent
where $\left[ \cdot \right]$ is the floor function, $g$ is the genus of $\Gamma_{0}(N) \backslash \Tilde{\mathbb{H}}$, $v_{\infty}$ is the number of inequivalent cusps of $\Gamma_{0}(N)$, $v_{2}$ and $v_{3}$ are the number of inequivalent fixed points of $\Gamma_{0}(N)$ of order 2 and 3, and $\delta_{k,2}=1$ if $k=2$ and otherwise $0$.

To calculate the dimensions of $M_{k} \left( \Gamma_{0}(N) \right)$ and $S_{k} \left( \Gamma_{0}(N) \right)$ explicitly, we have to calculate $g,v_{\infty},v_{2}$ and $v_{3}$. There are formulae to calculate them:

\noindent
\begin{align*}
    &g=1+\frac{1}{12}I - \frac{1}{4}v_{2} -\frac{1}{3}v_{3} -\frac{1}{2}v_{\infty}\\
    &v_{\infty} = \sum_{\{d|N\}} \phi \left( \operatorname{gcd} \left( d,\frac{N}{d} \right) \right),\\
    &v_{2}=\epsilon \left( \frac{N}{4} \right) \prod_{\{p|N\}} \left( 1+\left( \frac{-4}{p} \right)_{\mathrm{L}} \right),\\
    &v_{3}=\epsilon \left( \frac{N}{9} \right) \prod_{\{p|N\}} \left( 1+\left( \frac{-3}{p} \right)_{\mathrm{L}} \right),\\
\end{align*}
\noindent
where $\phi$ is the Euler function, $\epsilon(x)=1-\delta_{\mathbb{Z}}=1$ if $x \notin \mathbb{Z}$, $(\frac{\cdot}{\cdot})_{\mathrm{L}}$ is the Legendre symbol and $I$ is the index of $\Gamma_{0}(N)$ in $SL(2,\mathbb{Z})$. Moreover, $\{ d|N \}$ is a set of divisors of $N$ and $\{ p|N\}$ is a set of prime divisors of $N$.

\subsection{Dedekind $\eta$-function and eta quotients}

In the subsection~\ref{2.2}, we defined automorphic forms and cusp forms of weight $k$ with respect to $\Gamma_{0}(N)$. In this subsection, we introduce the Dedekind $\eta$-function and eta quotients as examples.

The Dedekind $\eta$-function is defined by

\noindent
\begin{align*}
    \eta(z) = q^{\frac{1}{24}} \prod_{n \geq 1} \left(1-q^{n} \right),
\end{align*}
\noindent
where $q=e^{2\pi i z}$.

If $\gamma=\begin{pmatrix}a & b \\ c & d\end{pmatrix} \in SL(2,\mathbb{Z})$, the Dedekind $\eta$-function satisfies (c.f., \cite{Knopp:1970})

\noindent
\begin{align}
\label{eta-trans}
    \eta \left( \gamma(z) \right) = v(\gamma) (cz+d)^{1/2} \eta(z),    
\end{align}
where
\begin{align*}
    v(\gamma)=
    \begin{cases}
    &\left( \frac{d}{|c|} \right)_{\mathrm{KL}} \exp \left( \frac{\pi i}{12} \left( (a+d-3)c-bd(c^{2}-1) \right) \right) \quad \text{if } 2 \nmid c, \\
    &\left( \frac{c}{|d|} \right)_{\mathrm{KL}} \exp \left( \frac{\pi i}{12} \left( (a-2d)c - bd(c^{2}-1) + 3d - 3 \right) \right) \epsilon(c,d) \quad \text{if } 2 \mid c,
    \end{cases}
\end{align*}
where $\left( \frac{\cdot}{\cdot} \right)_{\mathrm{KL}}$ is the Kronecker-Legendre symbol and $\epsilon(c,d)=-1$ if $c \leq 0$ and $d <0$ and $\epsilon(c,d)=1$ for otherwise.

Since $\Gamma_{0}(N) \subset SL(2,\mathbb{Z})$, the transformation law~(\ref{eta-trans}) is valid if $\gamma \in \Gamma_{0}(N)$. This means that the Dedekind $\eta$-function is a automorphic form of weight $\frac{1}{2}$ with the character $v$ with respect to $\Gamma_{0}(N)$.

Actually, automorphic forms or cusp forms of integer weight will be necessary. According to Shimura correspondence~\cite{Shimura:1973}, such a form can be constructed from automorphic forms of half integer weight like the Dedekind $\eta$-function.

Eta quotients are defined by

\noindent
\begin{align*}
    f(z)=\prod_{\{d|N\}} \eta^{r_{d}}(dz),
\end{align*}
where $N$ is a positive integer and $r_{d} \in \mathbb{Z}$.

The functions $f(z)=\prod_{\{d|N\}} \eta^{r_{d}}(dz)$ are likely to become automorphic forms of weight $\frac{1}{2} \sum_{\{d|N\}} r_{d}$, but in fact, there are some requirements.

\begin{theo}
\label{theorem-etaquo}
    (c.f.,~\cite{Newman:1957,Newman:1959,Ligozat:1975,Gordon:1989}) Let $f(z)=\prod_{\{d|N\}} \eta^{r_{d}}(dz)$ and $k=\frac{1}{2} \sum_{\{d|N\}} r_{d}$. The function $f$ is a automorphic form of weight $k$ with character $\chi$ with respect to $\Gamma_{0}(N)$ if and only if
    \begin{align}
        \label{cond1}
        &\sum_{\{d|N\}} dr_{d} \equiv 0 \quad (\operatorname{mod} 24),\\
        \label{cond2}
        &\sum_{\{d|N\}} \frac{N}{d}r_{d} \equiv 0 \quad (\operatorname{mod} 24),\\
        &P= \prod_{\{d|N\}} d^{r_{d}} \text{ is a rational square}, \notag
    \end{align}
\end{theo}
where the character $\chi(\gamma)=\left( \frac{(-1)^{k}P}{d} \right)_{\mathrm{L}}$ for $\gamma=\begin{pmatrix}
    a & b \\
    c & d
    \end{pmatrix}
\in \Gamma_{0}(N)$.
In the following, we denote $[r_{1}, \dots, r_{N}]_{k}=\prod_{\{d|N\}} \eta^{r_{d}}(dz)$.

If the level $N$ is fixed, the conditions~(\ref{cond1}) and~(\ref{cond2}) are Diophantine equations. Numerical calculation support us to systematically construct eta quotients, and we will consider in section~\ref{4}.

\section{Dirac operator on $\Gamma \backslash \Tilde{\mathbb{H}}$ in gauge background}
\label{3}

We interested in the zero mode solutions of a Dirac operator on a magnetized Riemann surface of higher genus. Such a Dirac operator $D$ is given by

\noindent
\begin{align}
\label{gen-dirac}
    D=i \sigma^{a} {\theta_{a}}^{\mu} \left( \partial_{\mu} + \frac{1}{4} \Omega_{\mu b c}\sigma^{b} \sigma^{c} - i M A_{\mu} \right),
\end{align}
where $\sigma^{a}\ (a=1,2)$ are Pauli matrices, $\Omega_{\mu a b}\ (\mu=1,2)$ are Spin connection, ${\theta_{a}}^{\mu}$ are the inverse of the zweibein for the metric, and $M$ is the charge of a field that the Dirac operator acts on. Moreover, we use the Roman character for the index of a flat space and the Greek character for that of a curved space\footnote{We denote $x^{1}=x,x^{2}=y$ and $z=x+iy$.}.

To obtain a Dirac operator~(\ref{gen-dirac}), in the following, we calculate the spin connection and select a $U(1)$ gauge background.

\subsection{Zweibein and spin connection}
\label{3.1}

Let us start from $\mathbb{H}$ and the metric~(\ref{metric}). Since the metric~(\ref{metric}) is invariant under the M$\Ddot{\text{o}}$bius transformations, we can adopt the metric~(\ref{metric}) of $\Gamma \backslash \mathbb{H}$. The zweibein ${e_{\alpha}}^{i}$ are defined by 

\noindent
\begin{align*}
    g_{\alpha \beta}={e_{\alpha}}^{a} {e_{\beta}}^{b} \eta_{ab}, \quad \eta_{ab}=\operatorname{diag}(+1,+1).
\end{align*}
From the metric~(\ref{metric}), the zweibein and its inverse are

\noindent
\begin{align}
\label{zwei}
    {e_{1}}^{1}={e_{2}}^{2}=\frac{1}{y}, \quad {\theta_{1}}^{1}={\theta_{2}}^{2}=y, \quad \text{otherwise}=0.
\end{align}

The spin connection (in the case of torsionfree) $\Omega_{\mu a b}$ is determined by

\noindent
\begin{align*}
    de^{a} + \Omega^{ab} \wedge e_{b}=0.
\end{align*}
By solving this equation, the spin connection is

\noindent
\begin{align}
\label{spinconnection}
    {\Omega_{1}}^{12}=-{\Omega_{1}}^{21}=-\frac{1}{y}, \quad \text{otherwise}=0.
\end{align}

\subsection{Selection of gauge background}
\label{3.2}

In this paper, we select the symplectic gauge potential as a $U(1)$ gauge background. The symplectic gauge potential is defined by the symplectic form. In the local coordinate, we can write the symplectic form $w$ as

\noindent
\begin{align*}
    w=i g_{z\bar{z}}dz \wedge d\bar{z}.
\end{align*}
From the metric~(\ref{metric}), the symplectic form is

\noindent
\begin{align*}
    w=\frac{1}{y^{2}} dx \wedge dy.
\end{align*}
Then, the symplectic gauge potential $A$ is defined by 

\noindent
\begin{align*}
    w = \frac{V}{\pi} dA.
\end{align*}
where $V$ is the volume of the base manifold and the prefactor is for convenience. In this paper, the base manifold is $\Gamma_{0}(N) \backslash \Tilde{\mathbb{H}}$ and its volume is given by $V\left( \Gamma_{0}(N) \backslash \Tilde{\mathbb{H}} \right) = \frac{\pi}{3} I$, where $I$ is the index of $\Gamma_{0}(N)$ in $SL(2,\mathbb{Z})$.
Explicitly, the symplectic gauge potential is given by

\noindent
\begin{align}
\label{gauge}
    A=\frac{iQ}{2} \frac{dz + d \bar{z}}{z-\bar{z}}, \quad Q:=\frac{2\pi}{V}.
\end{align}

we have to consider gauge transformations and a boundary condition. In the following, we select the twisted boundary condition. The twisted boundary condition represent the consistency between gauge transformations and the identification for the base manifold, e.g., gauge transformations and the shift operations to define a torus (c.f., Ref.~\cite{Cremades:2004wa}). In this paper, we have to consider the gauge transformations under $\Gamma_{0}(N)$.

Actually, the M$\ddot{\text{o}}$bius transformations~(\ref{Mobius}) induce the gauge transformations of the symplectic gauge potential~(\ref{gauge}) 

\noindent
\begin{align*}
    A \rightarrow A + d\lambda, \quad \lambda=\frac{iQ}{2} \log \left( \frac{c\bar{z}+d}{cz+d} \right).
\end{align*}
Accordingly, if we consider a charged scalar field $\phi(z,\bar{z})$ with the charge $M$, the gauge transformations of such a field are 

\noindent
\begin{align*}
    \phi(z,\bar{z}) \rightarrow \phi \left( \gamma(z),\gamma(\bar{z}) \right)= \left(  \frac{cz+d}{c\bar{z}+d} \right)^{MQ/2} \phi(z,\bar{z}).
\end{align*}
The twisted boundary condition is defined by restricting $SL(2,\mathbb{R})$ to $\Gamma_{0}(N)$.

\subsection{Dirac operator}
\label{3.3}

Based on the subsections~\ref{3.1} and \ref{3.2}, we can obtain the Dirac operator in the $U(1)$ gauge background. By substituting Eqs.~(\ref{zwei}),~(\ref{spinconnection}) and~(\ref{gauge}) to Eq.~(\ref{gen-dirac}), the Dirac operator for a charged fermion with the charge $M$ is given by

\noindent
\begin{align*}
    D =
    \begin{pmatrix}
    0 & (z-\bar{z})\partial_{z} + \frac{1}{2} (MQ -1) \\
    (z-\bar{z})\partial_{\bar{z}} + \frac{1}{2} (MQ +1)
    \end{pmatrix}.
\end{align*}
If we label the components of a fermion by the eigenvalue of the chirality matrix $\sigma_{3}$, i.e.,

\noindent
\begin{align*}
    \psi = \begin{pmatrix}
    \psi_{+1}\\
    \psi_{-1}
    \end{pmatrix},
\end{align*}
the zero mode equations can be written as

\noindent
\begin{align}
\label{zeroeq}
    \left[ (z - \bar{z})(\partial_{x}+is \partial_{y}) + \frac{1}{2}\left( NQ+s \right) \right] \psi_{s}=0 \quad (s=\pm1).
\end{align}
Since the spin connection plays a role like the symplectic gauge potential, the twisted boundary condition for a zero mode solution $\psi_{s}$ is written as

\noindent
\begin{align}
\label{f-gtrans}
    \psi_{s}( \gamma(z), \gamma(\bar{z})) = \left(   \frac{cz+d}{c\bar{z}+d} \right)^{(MQ+s)/2} \psi_{s}(z, \bar{z}), \quad ^{\forall}\gamma \in \Gamma_{0}(N).
\end{align}

The following ansatz can satisfy the zero mode equations~(\ref{zeroeq}):

\noindent
\begin{align*}
    \psi_{+}(z, \bar{z}) = (z-\bar{z})^{(1 + MQ)/2} h_{+}(z), \quad \psi_{-}(z, \bar{z}) = (z-\bar{z})^{(1 - MQ)/2} h_{-}(\bar{z}),
\end{align*}
where $h_{+}(z)$ and $h_{-}(\bar{z})$ are arbitrary holomorphic and anti-holomorphic functions, respectively. Because of the gauge transformations~(\ref{f-gtrans}), we can show that $h_{+}(z)$ and $h_{-}(\bar{z})$ must satisfy

\noindent
\begin{align}
\label{h}
    h_{+}\left( \gamma(z) \right) = (cz+d)^{1+MQ}h_{+}(z), \quad  h_{-}\left( \gamma(\bar{z}) \right) = (c\bar{z}+d)^{1-MQ}h_{-}(\bar{z}), \quad ^{\forall}\gamma \in \Gamma_{0}(N).
\end{align}
The condition~(\ref{h}) requires that $h_{+}(z)$ is an automorphic form of weight $1+MQ$ with respect to $\Gamma_{0}(N)$ ($h_{-}(\bar{z})$ must be an anti-holomorphic version of the automorpohic form of weight $1-MQ$). In addition, we have to consider following conditions.

\begin{itemize}
    \item Convergence
\end{itemize}
In this case, the normalization condition (the inner product) is given by, e.g.,

\noindent
\begin{align}
\label{inner}
    ||\psi_{+}||^{2}:= 2^{1+MQ}\int_{\Gamma_{0}(N) \backslash \Tilde{\mathbb{H}}} |h_{+}(x+iy)|^{2} y^{1+MQ} \frac{1}{y^{2}}dxdy < \infty.
\end{align}
This inner product is the Petersson inner product except for the overall factor. The convergence of the Petersson inner product is guaranteed for the cusp forms. Therefore, $h_{+}(z) \in S_{1+MQ} \left( \Gamma_{0}(N) \right)$ (Similarly, for $h_{-}(\bar{z})$). 

\begin{itemize}
    \item The behavior on the origin
\end{itemize}
The zero mode $\psi_{s}$ should be extended to the cusps of $\Gamma_{0}(N)$ lying $z-\bar{z}=0$. Therefore, $MQ \geq -1$ and $1 \geq  MQ$ for $\psi_{+}$ and $\psi_{-}$, respectively.

\begin{itemize}
    \item Selection of congruence subgroups
\end{itemize}
In this paper, we select $\Gamma_{0}(N)$ as a congruence subgroup. In general, $\Gamma_{0}(N)$ includes $-\mathbf{1}$. This fact restricts the weight of the automorphic forms must be a even integer since

\noindent
\begin{align*}
    f \in M_{k}(\Gamma_{0}(N)) \rightarrow f \left( -\mathbf{1}(z) \right) =f(z) = (-1)^{k} f(z).
\end{align*}
Therefore, 

\noindent
\begin{align*}
\begin{cases}
1+MQ \in 2\mathbb{Z}_{\geq 0} \geq 0 \quad \text{for } \psi_{+}, \\
1-MQ \in 2\mathbb{Z}_{\geq 0} \geq 0 \quad \text{for } \psi_{-}.
\end{cases}
\end{align*}
We can find that $MQ=0$ is excluded in the both cases. In addition, $MQ=-1$ for $\psi_{+}$ and $MQ=1$ for $\psi_{-}$ are also excluded since $\operatorname{dim}S_{0} \left( \Gamma_{0}(N) \right)=0$.

The result of the above three conditions implies that $h_{+}(z)$ ($h_{-}(\bar{z})$) is a cusp form and

\noindent
\begin{align*}
\begin{cases}
MQ \in 2\mathbb{Z}_{\geq 0} +1 \quad \text{for } \psi_{+}, \\
-MQ \in 2\mathbb{Z}_{\geq 0} +1 \quad \text{for } \psi_{-}.
\end{cases}
\end{align*}
Moreover, the chiral fermions can be realized since the requirement for $MQ$ separates completely. On the other hand, the number of the generations is given by the dimension formula~(\ref{casp-dimformula}) since the degree of freedom of the zero mode solutions is the selection of $h_{+}(z)$ or $h_{-}(\bar{z})$.

\section{Construction of zero mode solutions}
\label{4}

In the section~\ref{3}, we constructed the Dirac operator in the $U(1)$ gauge background on a magnetized Riemann manifold with higher genus, and the zero mode solutions are characterized by the cusp forms. It is sufficient to construct basis of $S_{k}(\Gamma_{0}(N))$ since $S_{k}(\Gamma_{0}(N))$ is a vector space on $\mathbb{C}$. To obtain the basis, we apply the method with eta quotients~\cite{Rouse:2013ar,Allen:2019ar}.

\subsection{Construction}
\label{4.1}
To show that an arbitrary base $f(z)$ can be written in eta quotients, we need to show e.g., $f(z)-\sum_{i} \alpha_{i} [r_{1,i}, \dots, r_{N,i}]=0$, where $\alpha_{i} \in \mathbb{C}$. Therefore, we apply following theorems to guarantee the equality. 

\begin{theo}
\label{order}
    (Theorem 1.65. in Ref.~\cite{Ono:2004}) Let $c,d$ and $N$ be positive integers with $d|N$ and $\operatorname{gcd}(c,d)=1$. If $f(z)$ is an eta quotient satisfying the conditions of Theorem~\ref{theorem-etaquo} for $N$, then the order of vanishing of $f(z)$ at the cusp $\frac{c}{d}$, which is denoted by $v_{\frac{c}{d}}$, is
    \begin{align*}
        v_{\frac{c}{d}} = \frac{N}{24} \sum_{\{\delta|N\}} \frac{\operatorname{gcd}^{2}(d,\delta) r_{\delta}}{\operatorname{gcd}\left( d,\frac{N}{d} \right) d \delta}.
    \end{align*}
\end{theo}
\noindent
The order $v_{\frac{c}{d}}$ relates to the other order $\operatorname{inv}_{\frac{d}{c}}$ that is the invariant order of vanishing at $\frac{c}{d}$. Actually, $v_{\frac{c}{d}}$ satisfies $v_{\frac{c}{d}} = h \cdot \operatorname{inv}_{\frac{d}{c}}$, where $h$ is the width of the cusp $\frac{c}{d}$.

\begin{theo}
\label{Sturm}
    (Sturm's bound: Theorem 6.4.7. in Ref.~\cite{Murty:2015}) Let $\Gamma$ be a congruence subgroup, which includes $\pm \mathbf{1}$ and $f \in M_{k}(\Gamma)$. Let $c_{1}, \dots, c_{t}$ be the $\Gamma$-inequivalent cusps of $\Gamma$. If
    \begin{align*}
        \sum_{i=1}^{t} \operatorname{inv}_{c_{i}}(f) > \frac{k}{12} [SL(2,\mathbb{Z}):\Gamma],
    \end{align*}
    then $f=0$, where $ [SL(2,\mathbb{Z}):\Gamma]$ is the index of $\Gamma$ in $SL(2,\mathbb{Z})$.
\end{theo}

\begin{theo}
\label{order-sum}
    (Theorem 4.2. and Remark 4.3. in Ref.~\cite{Allen:2019ar}) Let $N=p_{1} \cdots p_{t}$, a product of distinct primes with $p_{i} \geq 5$, and let $f(z)$ be a eta quotient satisfying the conditions of the theorem~\ref{theorem-etaquo}. Then,
    \begin{align*}
        \sum_{\{d|N\}} v_{\frac{1}{d}} = \frac{k}{12} \prod_{\{p|N\}} (p+1).
    \end{align*}
\end{theo}

From the theorem~\ref{Sturm}, two automorphic form $f,g \in M_{k} \left( \Gamma \right)$ are equal by considering their $q$-expansions if their $q$-expansions agree to a power of $q$ beyond the bound in theorem~\ref{Sturm}. The above theorems are valid for $\Gamma=\Gamma_{0}(N)$ and $f,g \in S_{k} \left( \Gamma_{0}(N) \right)$. In the method for constructing the basis of $ S_{k} \left( \Gamma_{0}(N) \right)$, the Sturm's bound will be used to confirm an equality.

In the following, we select the level $N=pq$, where $p$ and $q$ are distinct primes to apply the method mentioned in Ref~\cite{Allen:2019ar}. The cusps of $\Gamma_{0}$ $(N=pq)$ are $\{0, \frac{1}{p}, \frac{1}{q}, \infty \}$. In addition, the order $v_{\frac{1}{d}}$, where $d \in \{d|N\}$, is positive integer~\cite{Allen:2019ar}. 

From the above, we can construct basis of $S_{k}(\Gamma_{0}(N=pq))$ by the following steps:

\begin{enumerate}
\renewcommand{\labelenumi}{\arabic{enumi}).}
\item Fix a semiprime $N=pq$, where $p$ and $q$ are distinct primes.
\item Based on the theorem~\ref{order-sum}, obtain a list of all partitions of $\frac{k}{12}(p+1)(q+1)$ into four positive integers, i.e.,
    \begin{align*}
        v_{0}+v_{\frac{1}{p}} + v_{\frac{1}{q}} + v_{\infty} =\frac{k}{12}(p+1)(q+1), \quad v_{0},v_{\frac{1}{p}}, v_{\frac{1}{q}}, v_{\infty} \geq 1.
    \end{align*}
\item For each $(v_{0},v_{\frac{1}{p}}, v_{\frac{1}{q}}, v_{\infty})$, solve the equations
    \begin{align*}
    24v_{0}=N r_{1} + q r_{p} + p r_{q} + r_{N}, \\
    24v_{p}=q r_{1} + N r_{p} + r_{q} + p r_{N}, \\
    24v_{q}=p r_{1} + r_{p} + N r_{q} + q r_{N}, \\
    24v_{\infty}=r_{1} + p r_{p} + q r_{q} + N r_{N}.
    \end{align*}
    for unique solution $(r_{1},r_{p},r_{q},r_{N}) \in \mathbb{Q}^{4}$, and exclude $(r_{1},r_{p},r_{q},r_{N})$ if $(r_{1},r_{p},r_{q},r_{N}) \notin \mathbb{Z}^{4}$.
\item For each $(r_{1},r_{p},r_{q},r_{N})$, define a eta quotient, i.e.,
    \begin{align*}
        f(z)=\eta(z)^{r_{1}} \eta(pz)^{r_{p}} \eta(qz)^{r_{q}} \eta(Nz)^{r_{N}}=[r_{1},r_{p},r_{q},r_{N}]_{k},
    \end{align*}
    and confirm that $(r_{1},r_{p},r_{q},r_{N})$ satisfies the conditions of the theorem~\ref{theorem-etaquo}. Moreover, exclude $(r_{1},r_{p},r_{q},r_{N})$ if $[r_{1},r_{p},r_{q},r_{N}]_{k}$ is not a cusp form (the starting order of the q-expansion is $\mathcal{O}(q^{0})$).
\item Construct a maximally linear independent set, denoted by $(\text{eta quotients})_{k}$, of the eta quotients from the step 4.
\item Compare $\operatorname{dim}(\text{eta quotients})_{k}=:E_{N,k}$ and $\operatorname{dim}S_{k}\left(\Gamma_{0}(N)\right)=:D_{N,k}$. If $E_{N,k}=D_{N,k}$, the eta quotients of $(\text{eta quotients})_{k}$ are the basis of $S_{k}\left(\Gamma_{0}(N)\right)$. If $E_{N,k} < D_{N,k}$\footnote{In this paper, we search eta quotients that are cusp forms, i.e., $[r_{1},r_{p},r_{q},r_{N}]_{k} \in S_{k}\left(\Gamma_{0}(N)\right)$. Therefore, $(\text{eta quotients})_{k}  \subset S_{k}\left(\Gamma_{0}(N)\right)$, i.e., $E_{N,k} \leq D_{N,k}$.}, go to next step.
\item Repeat steps 2-6 for higher weight $k'>k$ until $S_{k'}\left(\Gamma_{0}(N)\right)$ has the basis of eta quotients, and $S_{k'-k}\left(\Gamma_{0}(N)\right)$ contains an eta quotient $[r_{1},r_{p},r_{q},r_{N}]_{k'-k}$
\item Generated missing basis of $S_{k}\left(\Gamma_{0}(N)\right)$ numerically, e.g., SageMath~\cite{sage:2020}.
\item Let $g(z)$ be a missing base of $S_{k}\left(\Gamma_{0}(N)\right)$. Write $g(z)[r_{1},r_{p},r_{q},r_{N}]_{k'-k}$ as a linear combination of the basis of $S_{k'}\left(\Gamma_{0}(N)\right)$ with Sturm's bound since $g(z)[r_{1},r_{p},r_{q},r_{N}]_{k'-k}$ must be in $S_{k'}\left(\Gamma_{0}(N)\right)$. Divide by $[r_{1},r_{p},r_{q},r_{N}]_{k'-k}$ and obtain the missing base $g(z)$.
\end{enumerate}

\subsection{Demonstration}
\label{4.2}
In the subsection~\ref{4.1}, we proposed the systematic construction for the basis of $S_{k}\left(\Gamma_{0}(N)\right)$. In this subsection, we demonstrate for a concrete case to confirm the validity of our construction. In the following, we select $\Gamma_{0}(35)$ as a congruence subgroup. The guiding principles for the selection of the level $N$ are the followings:

\begin{itemize}
    \item $N=pq$, where $p$ and $q$ are distinct primes more than $5$.
    \item $\operatorname{dim}S_{k} \left( \Gamma_{0}(N) \right) > 0$,
    \item Genus $g \geq 2$,
    \item $^{\exists}k$ such that $\operatorname{dim}S_{k} \left( \Gamma_{0}(N) \right) =3$
\end{itemize}
$N=35$ is the minimum level satisfying the above principles since $35=5 \times 7$, $g=3$, and $\operatorname{dim}S_{k} \left( \Gamma_{0}(N) \right)=4k-6+\delta_{k,2}$. The first two are necessary for our method to be valid. The third is because we are interested. The fourth is for phenomenological applications.

In the following, we consider $S_{2} \left( \Gamma_{0}(35) \right), S_{4} \left( \Gamma_{0}(35) \right)$ and $S_{6} \left( \Gamma_{0}(35) \right)$ as a demonstration. The highest Sturm's bound is $\mathcal{O}(q^{24})$ for $S_{6} \left( \Gamma_{0}(35) \right)$. Therefore, we will show the q-expansions up to $\mathcal{O}(q^{25})$ for convenience.

\begin{itemize}
    \item $S_{2} \left( \Gamma_{0}(35) \right)$
\end{itemize}
Let us start from $S_{2} \left( \Gamma_{0}(35) \right)$. From the dimension formula~(\ref{casp-dimformula}), $\operatorname{dim}S_{2} \left( \Gamma_{0}(35) \right)=3$. On the other hand, from the method, there are 3 eta quotients

\begin{table}[h]
    \centering
    \begin{tabular}{lll}
         $[0,2,2,0]_{2}$, & $[1,1,1,1]_{2}$, &$[2,0,0,2]_{2}$.
    \end{tabular}
\end{table}
We show their $q$-expansions in the Appendix~\ref{appA}, and they are linearly independent each other since the starting orders are different. Therefore, we can select them as the basis of $S_{2} \left( \Gamma_{0}(35) \right)$.

\begin{itemize}
    \item $S_{4} \left( \Gamma_{0}(35) \right)$
\end{itemize}
From the dimension formula~(\ref{casp-dimformula}), $\operatorname{dim}S_{4} \left( \Gamma_{0}(35) \right)=10$. However, there are 9 eta quotients, and 8 of them are linearly independent\footnote{The eta quotient $[0,0,4,4]_{4}$ can be written by the other eta quotients.}:

\begin{table}[h]
    \centering
    \begin{tabular}{llll}
         $[-1,5,5,-1]_{4}$, & $[0,4,4,0]_{4}$, &$[1,3,3,1]_{4}$, & $[2,2,2,2]_{4}$, \\
         $[3,1,1,3]_{4}$, & $[4,0,0,4]_{4}$, &$[4,4,0,0]_{4}$, & $[5,-1,-1,5]_{4}$. 
    \end{tabular}
\end{table}
\noindent
We show their $q$-expansions in the Appendix~\ref{appA}. Since there are the eta quotients whose stating order are $\mathcal{O}(q^{1}) \sim \mathcal{O}(q^{7})$ and $[4,4,0,0]$ and $[-1,5,5,-1]$ do not have $\mathcal{O}(q^{8})$, we need to prepare the basis of the starting order are $\mathcal{O}(q^{8})$ and $\mathcal{O}(q^{9})$ or $\mathcal{O}(q^{10})$ numerically. By SageMath~\cite{sage:2020}, the basis stating from $\mathcal{O}(q^{8})$ and $\mathcal{O}(q^{10})$ are obtained as

\noindent
\begin{align*}
    &f_{1}=q^{8} - 2q^{11} - 2q^{12} + 2q^{13} + q^{15} + 2q^{16} + 2q^{17} - 5q^{18} \\
    &\hspace{5cm}- 2q^{19} + 2q^{20} + 2q^{21} + q^{22} - 2q^{23} + 2q^{24} - 7q^{25} + O(q^{26}),\\
    &f_{2}=q^{10} - 2q^{11} - q^{12} + 3q^{14} + 2q^{15} - 2q^{16} + 3q^{17} - 8q^{18} \\
    &\hspace{5cm}+ q^{19} + 5q^{20} + q^{21} + 2q^{22} - 8q^{23} + 9q^{24} - 8q^{25} + O(q^{26}).
\end{align*}
The 8 eta quotients, $f_{1}$ and $f_{2}$ are linearly independent. However, $f_{1}$ and $f_{2}$ are not linear combinations of the eta quotients belonging to $S_{4} \left( \Gamma_{0}(35) \right)$. 

\begin{itemize}
    \item $S_{6} \left( \Gamma_{0}(35) \right)$
\end{itemize}
From the dimension formula~(\ref{casp-dimformula}), $\operatorname{dim}S_{4} \left( \Gamma_{0}(35) \right)=18$. In this case, there are 39 eta quotients, and 18 of them are linearly independent\footnote{The eta quotients $[0,2,0,10]_{6},[-1,3,7,3]_{6},[-1,7,1,5]_{6},[-2,8,8,-2]_{6},[1,1,-1,11]_{6},[0,2,6,4]_{6},$\\ $[0,6,0,6]_{6},[-1,7,7,-1]_{6},[2,0,-2,12]_{6},[1,1,5,5]_{6},[0,2,12,-2]_{6},[1,5,-1,7]_{6},[0,6,6,0]_{6},[2,0,4,6]_{6},$\\ $[1,1,11,-1]_{6},[3,-1,3,7]_{6},[2,0,10,0]_{6},[4,2,2,4]_{6},[5,1,1,5]_{6},[6,0,0,6]_{6},[5,1,7,-1]_{6}$ can be written by the other eta quotients.}:

\begin{table}[h]
    \centering
    \begin{tabular}{lllll}
        $[-2,12,2,0]_{6}$, & $ [0,10,0,2]_{6}$, & $ [1,5,5,1]_{6}$, & $[2,4,4,2]_{6}$, & $[3,3,3,3]_{6}$,\\
        $[3,7,3,-1]_{6}$, & $[4,6,2,0]_{6}$, & $[5,5,1,1]_{6}$, & $[6,0,6,0]_{6}$, & $[6,4,0,2]_{6}$,\\ $[7,-1,-1,7]_{6}$, & $[7,-1,-1,7]_{6}$, & $[7,-1,5,1]_{6}$, & $[7,3,-1,3]_{6}$, & $[8,-2,-2,8]_{6}$,\\
        $[10,0,2,0]_{6}$, & $[11,-1,1,1]_{6}$, & $[12,-2,0,2]_{6}$. 
    \end{tabular}
\end{table}
\noindent
We show their $q$-expansions in the Appendix~\ref{appA}. 

By considering the product of $f_{1}$ or $f_{2}$ and $[0,2,2,0]_{2},[1,1,1,1]_{2}$ or $[2,0,0,2]_{2}$, we can obtain $f_{1}$ and $f_{2}$ as the combinations of the eta quotients:

\noindent
\begin{align*}
    f_{1}&=\frac{1}{[0,2,2,0]_{2}} \sum_{i=1}^{18} \alpha_{1,i} [r_{1,i},r_{5,i},r_{7,i},r_{35,i}]_{6} \\
        &=\frac{1}{[1,1,1,1]_{2}} \sum_{i=1}^{18} \alpha_{2,i} [r_{1,i},r_{5,i},r_{7,i},r_{35,i}]_{6} \\
        &=\frac{1}{[2,0,0,2]_{2}} \sum_{i=1}^{18} \alpha_{3,i} [r_{1,i},r_{5,i},r_{7,i},r_{35,i}]_{6},\\
    f_{2}&=\frac{1}{[0,2,2,0]_{2}} \sum_{i=1}^{18} \beta_{1,i} [r_{1,i},r_{5,i},r_{7,i},r_{35,i}]_{6} \\
        &=\frac{1}{[1,1,1,1]_{2}} \sum_{i=1}^{18} \beta_{2,i} [r_{1,i},r_{5,i},r_{7,i},r_{35,i}]_{6} \\
        &=\frac{1}{[2,0,0,2]_{2}} \sum_{i=1}^{18} \beta_{3,i} [r_{1,i},r_{5,i},r_{7,i},r_{35,i}]_{6}.\\
\end{align*}
The coefficients $\alpha_{k,i},\beta_{k,i}$ $(k=1 \sim 3, i=1 \sim 18)$\footnote{The index $i$ is assigned in the order of the eta quotients which the q-expansion are shown, e.g., $i=1 \rightarrow [-2,12,2,0]_{6}$, $i=2 \rightarrow [-1,11,1,1]_{6}$.} are uniquely determined. They are listed in the table~\ref{t1} in the Appendix~\ref{appA}.

From the above, The validity of our method is confirmed by the demonstration. We note that the basis by the eta quotients are not orthogonal to each other. To obtain the orthonormal basis, we have to apply the Gram–Schmidt orthonormalization. Therefore, in principle, the orthonormal basis can be constructed. In the Gram-Schmidt orthonomalization, we have to calculate the Petersson inner products~(\ref{inner}). We may calculate them by the Nelson's formula~\cite{Nelson:2015}. However, it is actually difficult to calculate the inner product.

\section{Applications}
\label{5}

We constructed the zero mode solutions of the Dirac operator on a magnetized Riemann surface of higher genus, $\Gamma_{0}(N) \backslash \Tilde{\mathbb{H}}$. In this section, we discuss applications of the result, especially for extra dimensional models and matrix regularization.

\subsection{Extra dimensional models}
\label{5.1}

As we mentioned in the section~\ref{1}, the zero mode solutions contribute some parameters like Yukawa couplings in 4D effective theory. For Yukawa couplings, we have to consider bosonic fields and their Kaluza-Klein expansion with respect to a suitable Laplacian. In the following, we focus on scalar eigenfunctions since scalar eigenfunctions of the Laplace operator can generate vector eigenfunctions~\cite{Conlon:2008qi}.
On dimensional reduction, the scalar eigenfunctions $\phi$ with the charge $M'$ and the eigenvalue $m^{2}$ satify

\noindent
\begin{align*}
    -g^{\alpha \beta} D_{\alpha} D_{\beta} \phi = m^{2} \phi,
\end{align*}
where $D_{\alpha}=\partial_{\alpha} - iM'A_{\alpha}$ is the $U(1)$ gauge covariant derivative. From the selection of the gauge background~(\ref{gauge}), when we consider a scalar field $\phi$ with the charge $M'$,

\noindent
\begin{align}
\label{lap}
    -g^{\alpha \beta} D_{\alpha} D_{\beta} \phi &= -2g^{z \bar{z}} D_{\bar{z}} D_{z} \phi - g^{\bar{z} z} \left[D_{z},D_{\bar{z}} \right]\phi \notag \\
    &=-2g^{z \bar{z}} D_{\bar{z}}D_{z} \phi - \frac{M'Q}{2} \phi \notag \\
    &=\left[ -y^{2} \left( \partial_{x}^{2} + \partial_{y}^{2} \right) + iM'Qy \partial_{x} + \frac{(M'Q)^{2}}{4} \right] \phi \notag \\
    &=\left[ \Delta_{M'Q} + \frac{(M'Q)^{2}}{4} \right]\phi,
\end{align}
where $\Delta_{M'Q} := -y^{2} \left( \partial_{x}^{2} + \partial_{y}^{2} \right)+ iM'Qy \partial_{x}$ is the hyperbolic Laplacian of weight $M'Q$. The lowest mode has the tachyonic mass coming from the negative curvature of a Riemann surface of higher genus. Actually, the lowest eigenvalue of $\Delta_{M'Q}$ is $\frac{M'Q}{2}(1-\frac{M'Q}{2})$ (c.f., Ref.~\cite{Cohen:2017}). Such a mode can be obtained by solving $D_{z}\phi=0$ since the second line of eq.~(\ref{lap}) and $-2g^{z \bar{z}} D_{\bar{z}}D_{z}$ is positive semi-definite. Based on our discussion for fermions, the solution of $D_{z}\phi=0$ is in $S_{M'Q} \left( \Gamma_{0}(N) \right)$, and we have already constructed. Therefore, Yukawa couplings and more general n-point couplings can be calculated by the overlap integrals (c.f.,~\cite{Honda:2018sjy}) that are described by the Petersson inner product of some combinations of eta quotients. The formula by Ichino~\cite{Ichino:2008} may be useful, and the author of Ref.~\cite{Collins:2018} attempted to calculate the Ichino's formula from a numerical approach. However, it is still difficult to obtain explicit values.

\subsection{Matrix regularization}
\label{5.2}
Let $f$ be a function on a Riemann surface of higher genus, $\Gamma_{0}(N) \backslash \Tilde{\mathbb{H}}$. The matrix regularization of $f$ into a $n \times n$ matrix $T_{n}(f)$ is defined by

\noindent
\begin{align*}
    T_{n}(f)_{IJ}=\int_{\Gamma_{0}(N) \backslash \Tilde{\mathbb{H}}} w \psi^{\dagger}_{J} \cdot f \psi_{I},
\end{align*}
where $\cdot$ describes the contraction of spinor indices, $w$ is the volume form (same with the symplectic form since we consider a two dimensional manifold) and $\{\psi_{I} \mid I=1,2,\dots,n \}$ is an orthonomal basis of the zero mode solutions of a Dirac operator in a $U(1)$ gauge background. From our result, such basis can be generated by the eta quotients and their Gram–Schmidt orthonormalization. As a target of matrix regularization, we are interested in the basis of function space on $\Gamma_{0}(N) \backslash \Tilde{\mathbb{H}}$. They are obtained by the spectral decomposition of $L^{2}\left(\Gamma_{0}(N) \backslash \Tilde{\mathbb{H}} \right)$ with respect to the hyperbolic Laplacian of weight 0, $\Delta_{0}$, since typical target functions of matrix regularization are scalar fields with charge $0$. The eigenfunctions of $\Delta_{0}$ are called Maass forms, which transform like automorphic forms but not holomorphic functions. Since $\begin{pmatrix}1 & 1 \\ 0 & 1\end{pmatrix} \in \Gamma_{0}(N)$, the Maass forms $f_{\text{M}}$ should be

\noindent
\begin{align*}
    f_{\text{M}} = \sum_{n=1}^{\infty} a_{n} \sqrt{y} K_{iR}(2 \pi n y)e^{2 \pi i n x}, 
\end{align*}
where $K_{\alpha}(\cdot)$ is the second modified Bessel function with the parameter $\alpha$ and $R$ is the spectral parameter characterizing the eigenvalue of $\Delta_{0}$ as $\frac{1}{4}+R^{2}$. From the above, we expect that the target functions are $e^{2 \pi i  x}$ and $ \sqrt{y} K_{iR}(2 \pi  y)$ if we refer the matrix regularization on $T^{2}$ (c.f., Ref.~\cite{Adachi:2020asg}). The spectral parameter $R$ may play a role of a quantum number, c.f., the orbit angular momentum for spherical harmonics for matrix regularization on $S^{2}$.

\section{Conclusion and discussion}
\label{6}

In this paper, we have studied the zero mode system of a Dirac operator on a magnetized Riemann surface of higher genus as the quotient $\Gamma_{0}(N) \backslash \Tilde{\mathbb{H}}$. In section~\ref{3}, we mentioned the realization of chiral fermions. In addition, we note that the zero mode solutions should be cusp forms $S_{k} \left( \Gamma_{0}(N) \right)$. In section~\ref{4}, we presented a method to construct the basis of $S_{k} \left(\Gamma_{0}(N) \right)$ by eta quotients and restricting the level $N$. To confirm the method, we demonstrated for $S_{2} \left( \Gamma_{0}(35) \right),S_{4} \left( \Gamma_{0}(35) \right)$ and $S_{6} \left( \Gamma_{0}(35) \right)$ in subsection~\ref{4.2}. In section~\ref{5}, we discussed extra dimensional models and matrix regularization as applications of our result. 

We mentioned Yukawa couplings in section~\ref{5.1}. On the other hand, in the context of Modular flavor symmetry (e.g., Ref.~\cite{Feruglio:2017spp}), Yukawa couplings are described by linear combinations of the Dedekind $\eta$-function and its derivative. Therefore, our method and result may be compatible with such a context.

In this paper, we selected $\Gamma_{0}(35)$ as an example. In fact, it is important to be able to generate enough eta quotients. In that sense, it may be possible to relax the guiding principles in the subsection~\ref{4.2}. For example, $\Gamma_{0}(30)$\footnote{$\Gamma_{0}(30) \backslash \Tilde{\mathbb{H}}$ has the same genus, $g=3$, with $\Gamma_{0}(35) \backslash \Tilde{\mathbb{H}}$.} satisfies the guiding principles, except the first one. We may extend the scope of our method since there are several works for such a direction (e.g., Ref.~\cite{Rouse:2013ar}).

On the other hand, quotient manifolds $\Gamma_{0}(N) \backslash \Tilde{\mathbb{H}}$ include surfaces with genus 0 and 1. Our method can be extended to such cases if enough eta quotients can be generated. Zero mode solutions for magnetized sphere and magnetized torus have been already known~\cite{Conlon:2008qi,Cremades:2004wa}. Especially in the case of magnetized torus, there are several phenomenological discussions (e.g., Ref.~\cite{Abe:2012fj}). We are also interested in the difference between them and our model.

\section*{Acknowledgments}
The author would like to thank H. Abe for helpful comments.

\def\thesubsection{\Alph{subsection}}
\setcounter{subsection}{0}

\renewcommand{\theequation}{A.\arabic{equation}}
\setcounter{equation}{0}

\section*{Appendix}
\label{app}

\subsection{Explicit $q$-expansions and coefficients}
\label{appA}
In the following, we show the $q$-expansions of the linear independent eta quotients in $S_{2} \left( \Gamma_{0}(35) \right), S_{4} \left( \Gamma_{0}(35) \right)$.

\begin{itemize}
    \item $S_{2} \left( \Gamma_{0}(35) \right)$
\end{itemize}

\noindent
\begin{align*}
    &[0,2,2,0]_{2}\\
    &=q-2 q^6-2 q^8-q^{11}+4 q^{13}-q^{15}+2 q^{16} \\
    &\hspace{6cm}+2 q^{18}+2 q^{20}+q^{21}+2 q^{22}-4 q^{23}+q^{25}+O\left(q^{26}\right),\\
    &[1,1,1,1]_{2}\\
    &=q^2-q^3-q^4+q^8+q^9+q^{10}+q^{11}-2 q^{12}+q^{13}-q^{14}-q^{15}-3 q^{16} \\
    &\hspace{6cm}-q^{17}+q^{18}+2 q^{19}-q^{20}+q^{21}-2 q^{23}+4 q^{24}+O\left(q^{26}\right),
\end{align*}
\begin{align*}
    &[2,0,0,2]_{2}\\
    &=q^3-2 q^4-q^5+2 q^6+q^7+2 q^8-2 q^9-2 q^{11}-2 q^{12}+q^{13}+2 q^{16} \\
    &\hspace{7cm}+3 q^{17}-2 q^{18}+2 q^{19}-2 q^{22}-2 q^{23}+O\left(q^{26}\right).
\end{align*}

\begin{itemize}
    \item $S_{4} \left( \Gamma_{0}(35) \right)$
\end{itemize}

\noindent
\begin{align*}
    &[-1,5,5,-1]_{4}\\
    &=q+q^2+2 q^3+3 q^4+5 q^5+2 q^6+6 q^7+2 q^9-5 q^{10}-3 q^{11}-19 q^{12}+2 q^{13}-24 q^{14}\\
    &-10 q^{15}-19 q^{16}+q^{17}-38 q^{18}+15 q^{19}-35 q^{20}+12 q^{21}-3q^{22}+12 q^{23} \\
    &-25 q^{24}+80 q^{25}+O\left(q^{26}\right), \\
    &[0,4,4,0]_{4}\\
    &=q^2-4 q^7-4 q^9+2 q^{12}+16 q^{14}+2 q^{16}+8 q^{17}-8 q^{19}-8 q^{21}-5 q^{22}+8 q^{23} -32 q^{24}+O\left(q^{26}\right), \\
    &[1,3,3,1]_{4}\\
    &=q^3-q^4-q^5-2 q^8+3 q^9+q^{10}+3 q^{11}+3 q^{12}-3 q^{13}+2 q^{15}-9 q^{16}-12 q^{17}+4 q^{18} \\
    &-5 q^{19}+7 q^{20}+12 q^{22}+8 q^{23}+5 q^{24}-11 q^{25}+O\left(q^{26}\right), \\
    &[2,2,2,2]_{4}\\
    &=q^4-2 q^5-q^6+2 q^7+q^8+2 q^{10}-2 q^{12}-2 q^{13}-8 q^{14}+4 q^{15}+q^{16}+q^{18}+10 q^{19}\\
    &+9 q^{20}+4 q^{21}-11 q^{22}-6 q^{23}-4 q^{24}-22 q^{25}+O\left(q^{26}\right), \\
    &[3,1,1,3]_{4}\\
    &=q^5-3 q^6+5 q^8-q^{10}-4 q^{11}-q^{12}-2 q^{13}+3 q^{15}+10 q^{16}+q^{17}-q^{18}-q^{19}-12 q^{20} \\
    &+7 q^{21}-13 q^{22}-10 q^{23}+q^{24}+11 q^{25}+O\left(q^{26}\right), \\
    &[4,0,0,4]_{4}\\
    &=q^6-4 q^7+2 q^8+8 q^9-5 q^{10}-4 q^{11}-10 q^{12}+8 q^{13}+9 q^{14}+14 q^{16} -16 q^{17}-10 q^{18}\\
    &-4 q^{19}-8 q^{21}+14 q^{22}+20 q^{23}+2 q^{24}+O\left(q^{26}\right), \\
    &[4,4,0,0]_{4}\\
    &=q-4 q^2+2 q^3+8 q^4-5 q^5-8 q^6+6 q^7-23 q^9+20 q^{10}+32 q^{11}+16 q^{12}-38 q^{13} \\
    &-24 q^{14}-10 q^{15}-64 q^{16}+26 q^{17}+92 q^{18}+100 q^{19} -40 q^{20}+12q^{21}-128 q^{22} \\
    &-78 q^{23}+25 q^{25}+O\left(q^{26}\right),
\end{align*}

\begin{align*}
    &[5,-1,-1,5]_{4} \\
    &=q^{7}-5 q^{8}+5 q^{9}+10 q^{10}-15 q^{11}-5 q^{12}-10 q^{13}+31 q^{14}+20 q^{15}-30 q^{16}+15 q^{17}-75 q^{18} \\
    &+25 q^{19}+50 q^{20}+2 q^{21}+25 q^{22}-80 q^{23}+105q^{24}-55 q^{25}+O\left(q^{26}\right).
\end{align*}

\begin{itemize}
    \item $S_{6} \left( \Gamma_{0}(35) \right)$
\end{itemize}

\begin{align*}
 &[-2,12,2,0]_{6}\\
 &=q^3+2 q^4+5 q^5+10 q^6+20 q^7+24 q^8+41 q^9+48 q^{10}+61 q^{11}+50 q^{12}+83 q^{13}\\
 &+40 q^{14}+67 q^{15}+8 q^{16}+44 q^{17}-92 q^{18}+7 q^{19}-226 q^{20}-80q^{21}-304 q^{22}-144 q^{23}\\
 &-510 q^{24}-36 q^{25}+O\left(q^{26}\right),\\
 &[-1,11,1,1]_{6}\\
 &=q^4+q^5+2 q^6+3 q^7+5 q^8-4 q^9-8 q^{11}-12 q^{12}-27 q^{13}+6 q^{14}-26 q^{15}+4 q^{16}-9 q^{17} \\
 &+31 q^{18}-23 q^{19}+67 q^{20}-12 q^{21}+93 q^{22}+54q^{23}+53 q^{24}-18 q^{25}+ O\left(q^{26}\right), \\
 &[0,10,0,2]_{6}\\
 &=q^5-10 q^{10}+35 q^{15}-30 q^{20}-105 q^{25}+O\left(q^{26}\right), \\
 &[1,5,5,1]_{6}\\
 &=q^4-q^5-q^6-4 q^9+5 q^{10}+q^{11}+5 q^{12}+5 q^{13}-5 q^{15}+9 q^{16}-25 q^{17}-25 q^{18}+9 q^{19} \\
 &-15 q^{20}+25 q^{22}+35 q^{23}+25 q^{24}+55 q^{25}+O\left(q^{26}\right), \\
 &[2,4,4,2]_{6}\\
 &=q^5-2 q^6-q^7+2 q^8+q^9-2 q^{10}+6 q^{11}-2 q^{13}-2 q^{14}-13 q^{15}+6 q^{17}-10 q^{18}-q^{19} \\
 &+42 q^{20}+4 q^{21}+16 q^{22}-10 q^{23}-6 q^{24}-25 q^{25}+O\left(q^{26}\right), \\
 &[3,3,3,3]_{6}\\
 &=q^6-3 q^7+5 q^9-3 q^{11}+2 q^{12}-3 q^{13}-6 q^{14}-6 q^{16}+21 q^{17}+9 q^{18}-6 q^{19}+12 q^{21} \\
 &-15 q^{22}-27 q^{23}-38 q^{24}+O\left(q^{26}\right),
\end{align*}
\begin{align*}
 &[3,7,3,-1]_{6}\\
 &=q-3 q^2+5 q^4-7 q^6+14 q^7-3 q^8-26 q^9+8 q^{11}+7 q^{12}+21 q^{13}+28 q^{14}+38 q^{16} \\
 &-119 q^{17}-69 q^{18}+14 q^{19}-56 q^{21}+116 q^{22}+186 q^{23}+112q^{24}+25 q^{25}+O\left(q^{26}\right), \\
 &[4,6,2,0]_{6}\\
 &=q^2-4 q^3+2 q^4+8 q^5-5 q^6-10 q^7+14 q^8-6 q^9-31 q^{10}+26 q^{11}+31 q^{12}+18 q^{13} \\
 &-20 q^{14}-14 q^{15}-38 q^{16}-36 q^{17}-82 q^{18}+70 q^{19}+176 q^{20}+40q^{21}+51 q^{22}-92 q^{23} \\
 &+45 q^{24}-200 q^{25}+O\left(q^{26}\right), \\
 &[5,5,1,1]_{6}\\
 &=q^3-5 q^4+5 q^5+10 q^6-15 q^7-11 q^8+20 q^9-q^{10}-30 q^{11}+50 q^{12}+34 q^{13}-30 q^{14} \\
 &-94 q^{15}-20 q^{16}+44 q^{17}-15 q^{18}+105 q^{19}+166 q^{20}+60q^{21}-269 q^{22}-130 q^{23} \\
 &-265 q^{24}+90 q^{25}+O\left(q^{26}\right), \\
 &[6,0,6,0]_{6}\\
 &=q^2-6 q^3+9 q^4+10 q^5-30 q^6+11 q^8+36 q^9+36 q^{10}-124 q^{11}-42 q^{12}+126 q^{13} \\
 &+49 q^{14}+24 q^{15}-243 q^{16}-76 q^{17}+441 q^{18}-18 q^{19}-56 q^{20}-294q^{21}\\
 &-360 q^{22}+568 q^{23}-6 q^{24}-180 q^{25}+O\left(q^{26}\right), \\
 &[6,4,0,2]_{6}\\
 &=q^4-6 q^5+9 q^6+10 q^7-30 q^8-4 q^9+35 q^{10}+6 q^{11}-40 q^{12}+50 q^{13}+20 q^{14} \\
 &-110 q^{15}-101 q^{16}+110 q^{17}+220 q^{18}-86 q^{19}+130 q^{20}-40q^{21}-390 q^{22} \\
 &-380 q^{23}+200 q^{24}+430 q^{25}+O\left(q^{26}\right), \\
 &[7,-1,-1,7]_{6}\\
 &=q^{10}-7 q^{11}+14 q^{12}+7 q^{13}-49 q^{14}+22 q^{15}+28 q^{16}+56 q^{17}-49 q^{18}-168 q^{19}+128 q^{20} \\
 &-49 q^{21}+217 q^{22}+105 q^{23}-350 q^{24}-18q^{25}+O\left(q^{26}\right), 
 \end{align*}
 \begin{align*}
 &[7,-1,5,1]_{6}\\
 &=q^3-7 q^4+14 q^5+7 q^6-49 q^7+22 q^8+28 q^9+50 q^{10}-7 q^{11}-252 q^{12}+86 q^{13} \\
 &+245 q^{14}+85 q^{15}-63 q^{16}-677 q^{17}+213 q^{18}+882 q^{19}-180q^{20}-245 q^{21}-1307 q^{22} \\
 &+173 q^{23}+2320 q^{24}-305 q^{25}+O\left(q^{26}\right), \\
 &[7,3,-1,3]_{6}\\
 &=q^5-7 q^6+14 q^7+7 q^8-49 q^9+18 q^{10}+56 q^{11}-77 q^{13}+28 q^{14}+42 q^{15}-175 q^{16}\\
 &+21 q^{17}+315 q^{18}+224 q^{19}-478 q^{20}-56 q^{21}-119 q^{22}-560q^{23}+469 q^{24}+1050 q^{25}+O\left(q^{26}\right),\\
 &[8,-2,-2,8]_{6}\\
 &=q^{11}-8 q^{12}+20 q^{13}-70 q^{15}+66 q^{16}+40 q^{17}+42 q^{18}-141 q^{19}-260 q^{20}+441 q^{21}\\
 &-68 q^{22}+342 q^{23}-170 q^{24}-1105 q^{25}+O\left(q^{26}\right), \\
 &[10,0,2,0]_{6}\\
 &=q-10 q^2+35 q^3-30 q^4-105 q^5+238 q^6-262 q^8-145 q^9+70 q^{10}+1114 q^{11}-560 q^{12} \\
 &-1071 q^{13}-196 q^{15}+2502 q^{16}+140 q^{17}-2078 q^{18}-735 q^{19}-868q^{20}+2401 q^{21} \\
 &+1012 q^{22}-2684 q^{23}+2100 q^{24}+501 q^{25}-1638 +O\left(q^{26}\right),\\
 &[11,-1,1,1]_{6}\\
 &=q^2-11 q^3+44 q^4-55 q^5-110 q^6+375 q^7-154 q^8-419 q^9+11 q^{10}+341 q^{11}+1718 q^{12} \\
 &-2124 q^{13}-1651 q^{14}+1309 q^{15}+417 q^{16}+5844 q^{17}-4324q^{18}-5733 q^{19}-76 q^{20}\\
 &+901 q^{21}+15185 q^{22}-7372 q^{23}-10091 q^{24}+3510 q^{25}+O\left(q^{26}\right), \\
 &[12,-2,0,2]_{6}\\
 &=q^3-12 q^4+54 q^5-88 q^6-99 q^7+542 q^8-442 q^9-540 q^{10}+418 q^{11}+638 q^{12} \\
 &+2141 q^{13}-5000 q^{14}-1235 q^{15}+4852 q^{16}+583 q^{17}+9078 q^{18}-16898q^{19}-7870 q^{20} \\
 &+10000 q^{21}+7018 q^{22}+36438 q^{23}-50000 q^{24}-21295 q^{25}+O\left(q^{26}\right).
\end{align*}

\newpage
\begin{table}[h]
\begin{center}
{\renewcommand\arraystretch{1.5}
  \begin{tabular}{|c|c|c|c|c|c|c|} \hline
        & $\alpha_{1,i}$ & $\alpha_{2,i}$ & $\alpha_{3,i}$ & $\beta_{1,i}$ & $\beta_{2,i}$ & $\beta_{3,i}$ \\ \hline
        i=1 &$0$ &$0$ &$0$ &$0$ &$0$ &$0$ \\ \hline
        2 &$\frac{5}{343}$ &$0$ &$0$ &$-\frac{10}{2401}$ &$0$ &$0$ \\ \hline
        3 &$0$ & $\frac{5}{343}$&$0$ &$\frac{100}{2401}$ &$-\frac{10}{2401}$ &$0$ \\ \hline
        4 &$-\frac{16}{175}$ &$0$ &$\frac{1}{175}$ &$-\frac{43}{1225}$ &$\frac{4}{245}$ &$-\frac{2}{1225}$ \\ \hline
        5 &$\frac{22}{175}$ &$-\frac{16}{175}$ &$-\frac{2}{175}$ &$\frac{66}{1225}$ &$-\frac{83}{1225}$ &$\frac{24}{1225}$ \\ \hline
        6 &$\frac{7}{25}$ &$\frac{22}{175}$ &$-\frac{19}{175}$ &$\frac{157}{1225}$ &$\frac{6}{1225}$ &$-\frac{11}{175}$ \\ \hline
        7 &$\frac{94}{8575}$ &$\frac{1}{175}$ &$\frac{16}{8575}$ &$\frac{292}{60025}$ &$\frac{37}{60025}$ &$\frac{18}{60025}$ \\ \hline
        8 &$-\frac{27}{1125}$ &$-\frac{151}{8575}$ &$-\frac{51}{8575}$ &$-\frac{61}{8575}$ &$-\frac{293}{60025}$ &$-\frac{13}{60025}$ \\ \hline
        9 &$-\frac{468}{8575}$ &$-\frac{27}{1225}$ &$-\frac{117}{8575}$ &$-\frac{2049}{60025}$ &$\frac{253}{60025}$ &$-\frac{361}{60025}$ \\ \hline
        10 &$\frac{1156}{8575}$ &$\frac{303}{4900}$ &$\frac{183}{4900}$ &$\frac{15987}{240100}$ &$\frac{16}{8575}$ &$\frac{439}{34300}$ \\ \hline
        11 &$-\frac{311}{490}$ &$-\frac{1691}{4900}$ &$-\frac{43}{140}$ &$-\frac{1831}{6860}$ &$-\frac{2799}{17150}$ &$\frac{13}{6860}$ \\ \hline
        12 &$-\frac{169}{2450}$ &$-\frac{194}{6125}$ &$-\frac{1}{70}$ &$-\frac{577}{17150}$ &$-\frac{289}{85750}$ &$-\frac{31}{8575}$ \\ \hline
        13 &$-\frac{377}{2450}$ &$-\frac{1391}{17150}$ &$-\frac{104}{1225}$ &$-\frac{3956}{60025}$ &$-\frac{3973}{120050}$ &$-\frac{449}{17150}$ \\ \hline
        14 &$-\frac{377}{2450}$ &$-\frac{199}{2450}$ &$-\frac{104}{1225}$ &$-\frac{568}{8575}$ &$-\frac{81}{2450}$ &$-\frac{449}{17150}$ \\ \hline
        15 &$\frac{1083}{12250}$ &$\frac{1489}{24500}$ &$\frac{1119}{24500}$ &$\frac{7813}{171500}$ &$\frac{243}{12250}$ &$\frac{2097}{171500}$ \\ \hline
        16 &$-\frac{94}{8575}$ &$-\frac{1}{175}$ &$-\frac{16}{8575}$ &$-\frac{292}{60025}$ &$-\frac{37}{60025}$ &$-\frac{18}{60025}$ \\ \hline
        17 &$\frac{1}{70}$ &$\frac{398}{42875}$ &$\frac{123}{17150}$ &$\frac{23}{3430}$ &$\frac{2363}{600250}$ &$\frac{104}{60025}$ \\ \hline
        18 &$\frac{212}{42875}$ &$\frac{71}{24500}$ &$\frac{457}{171500}$ &$\frac{2739}{1200500}$ &$\frac{324}{300125}$ &$\frac{941}{1200500}$ \\ \hline
  \end{tabular}
  }
  \end{center}
  \caption{The coefficients of linear combinations with respect to the eta quotients}
  \label{t1}
\end{table}

\bibliographystyle{prsty}

\end{document}